%% file: Treiber_datafusion_small.tex
\documentclass[graybox]{svmult}

\usepackage{mathptmx}       % selects Times Roman as basic font
\usepackage{helvet}         % selects Helvetica as sans-serif font
\usepackage{courier}        % selects Courier as typewriter font
\usepackage{type1cm}        % activate if the above 3 fonts are
\usepackage{makeidx}         % allows index generation
\usepackage{graphicx}        % standard LaTeX graphics tool
\usepackage{multicol}        % used for the two-column index
\usepackage[bottom]{footmisc}% places footnotes at page bottom
\usepackage{caption}
\usepackage{multirow}
\usepackage{tabularx}
\usepackage{url}
\input{defsSubmit}
\makeindex             % used for the subject index
\begin{document}

\title*{Measuring and Modelling Crowd Flows - Fusing Stationary and
  Tracking Data}
\titlerunning{Measuring and Modelling Crowd Flows}% for an abbreviated version of
% your contribution title if the original one is too long
\author{Martin Treiber}
% Use \authorrunning{Short Title} for an abbreviated version of
% your contribution title if the original one is too long
\institute{Martin Treiber \at Technische Universit{\"a}t Dresden, W{\"u}rzburger Str. 35, D-01062, Dresden \email{treiber@vwi.tu-dresden.de}}
%
%!! falls Fehlermeldung at, altes svmult-> ersetzen!
% Use the package "url.sty" to avoid
% problems with special characters
% used in your e-mail or web address
%
\maketitle

\newcommand{\mt}[1]{#1} % use this to neutralize the text marking  

\abstract{The two main data categories of vehicular traffic flow,
  stationary detector data and floating-car data,
 are also available for many Marathons and other mass-sports
events: Loop detectors and other stationary data sources
find their counterpart in the RFID tags of the athletes recording the
split times at several stations during the race. Additionally, 
more and more athletes use smart-phone apps generating
track data points that are the equivalent of floating-car data.
We present a methodology to detect congestions  and 
estimate the location of jam-fronts, the delay
times, and the
spatio-temporal speed and density distribution of the athlete's crowd flow by
fusing these two data sources based on a first-order macroscopic model
with triangular fundamental diagram. The method can be used in
real-time or for analyzing past events. Using synthetic
``ground truth'' data generated by simulations with the
Intelligent-Driver Model, we show that, in a real-time application,
 the proposed algorithm is
robust and effective with minimal data requirements. Generally, two stationary data
sources and about ten ``floating-athlete'' trajectories per hour are sufficient.
%We apply this method to several Marathons including the Vasaloppet
%cross-country ski race 2014 and the Tel-Aviv
%Marathon 2014 and use the results to validate the macroscopic model. 
}

\section{\label{sec:1}Introduction}
While vehicular traffic data analysis and flow modeling is
a mature research field~\cite{TreiberKesting-Book}, only few
scientific investigations exist for the dynamics and data analysis of unidirectional
crowd flow, particularly in mass-sports events. 
Popular mass-sports events include classical Marathons (e.g., the \emph{New
York Marathon}), 
cross-country events (e.g.,
the \emph{Vasaloppet}~\cite{TGF13-ski}), and other events
such as the increasingly popular inline-skating nights (e.g., the \emph{Dresdner
  Nachtskaten}).

Microscopically, the crowd dynamics can be described either by  two-dimensional
active-particle systems~\cite{Helbing-01aa}, or by multi-lane
one-dimensional traffic flow models for the case of ski Marathons in
the classic style~\cite{TGF13-ski}. Unlike the situation in
general pedestrian 
traffic, the flow is (i) unidirectional and (ii) there is no route choice: All
athletes remain on the race track from start to finish. This means, the
dynamics is equivalent to that of mixed unidirectional
vehicular traffic flow on a link~\cite{Arasan-mixedTraffic,kanagaraj2013evaluation}.
%which is relevant for traffic flow in many developing

This allows to simplify the mathematical
description to a macroscopic, one-dimensional model. The free regime
is described in~\cite{TGF13-ski}. In the congested regime, most of the
individuality is lost. Moreover, genuine ``crowd-flow''
instabilities (stop-and go waves) are rare allowing to describe the macroscopic dynamics 
by Lighthill-Whitham-Richards (LWR) models~\cite{Lighthill-W}.

%#############################################################
\begin{figure}
\includegraphics[width=0.98\textwidth]{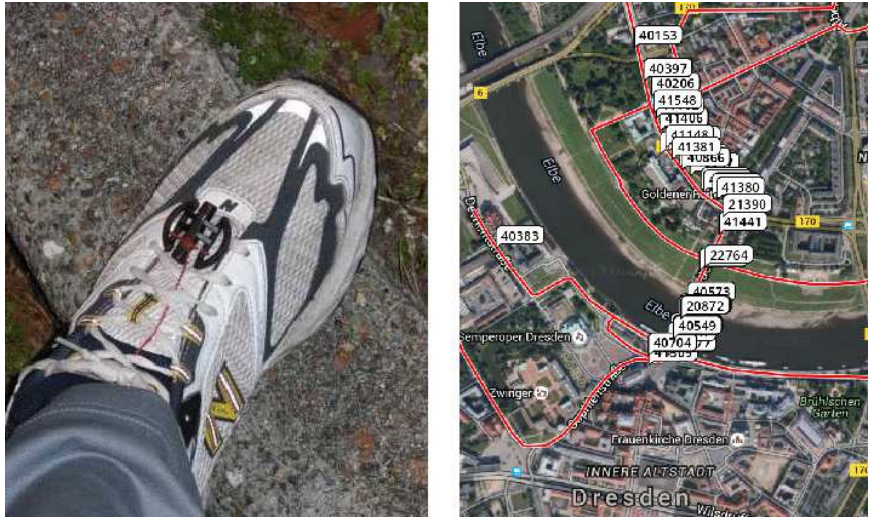}
\caption{\label{fig:data}Data basis for Marathon events:
  Stationary counting data at the stations obtained by RFID chips
  (left), and ``floating athlete data'' obtained by smart-phone apps
  (right).
}
\end{figure}
%#############################################################

These models can be calibrated by data that are essentially equivalent
to that of vehicular traffic.
Loop detectors and other stationary data sources
find their counterpart in the RFID tags of the athletes recording the
split times at several stations during the race
(cf. Fig.~\ref{fig:data} left). For our purposes,
they are simply counting detectors. Additionally, 
more and more athletes use smart-phone apps generating, in real time,
track data points that are the equivalent of floating-car data
(cf. Fig.~\ref{fig:data} right). 
These two data sources complement each other: split-time data are available
for all the athletes but only at a few positions. In contrast,
``floating-athlete data'' cover the whole track but the percentage of athletes
with activated smart-phone apps is low and unknown.

In this contribution, we present a methodology to detect congestions,
track their upstream fronts in real-time, and reconstruct the
spatio-temporal local speed  of the  crowd flow by
fusing these two data sources. The algorithm is based on a LWR model with
tridiagonal fundamental diagram. We calibrate and validate the method based on
simulated ``ground truth'' data generated with the 
Intelligent-Driver Model (IDM)~\cite{Opus}. 

In the next section, we develop the
estimation methodology, in Sect.~\ref{sec:res} we present the
simulation results before we conclude this paper with a discussion
(Sect.~\ref{sec:concl}).

%#############################################################
\section{\label{sec:meth}Methods}
%#############################################################

%#############################################################
\subsection{\label{sec:front}The reconstruction algorithm}
%#############################################################

\begin{figure}
\fig{01.00\textwidth}{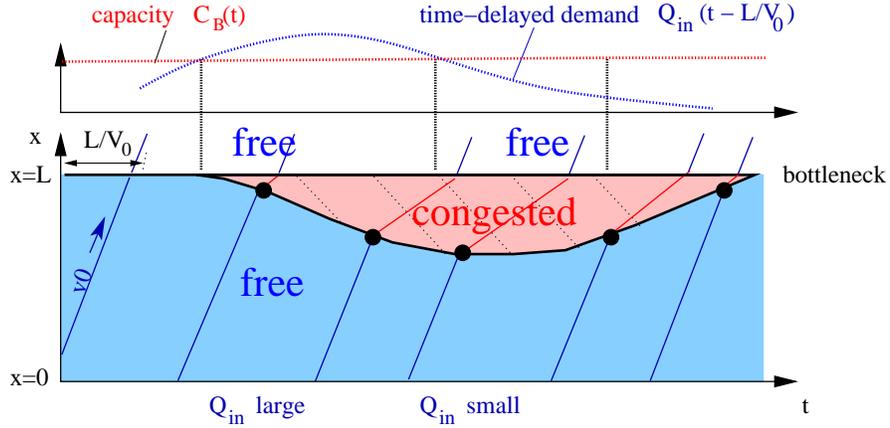}
\caption{\label{fig:shock}Schematic visualization of the basic
  reconstruction algorithm of Sect.~\ref{sec:front}: The flow
  information $Q\sub{in}$ of the upstream detector 
  at $x=0$ (demand) propagates downstream through free traffic at
  velocity $V_0$ (the density of the lines is proportional to
  $Q\sub{in}$) while the information about
  the bottleneck capacity $C\sub{B}$ (supply) propagates upstream
  through the congested traffic zone at velocity $c$ (dotted
  lines). At the transition, the shock-wave formula applies. 
  Shown are also some \emph{floating-athlete} trajectories. 
}
\end{figure}

The basic algorithm uses only the flow information of two stationary detectors that
are located at the upstream and downstream boundary of the section to
be analyzed. It is based on the LWR with the triangular fundamental diagram
$Q\sub{e}(\rho)$ which can be expressed in
terms of the parameters free-flow speed $V_0$, maximum flow (capacity)
$Q\sub{max}$, and wave speed $c$,
\be
\label{triang}
Q\sub{e}(\rho)= \left\{ \begin{array}{lll}
 V_0\rho & \quad \text{if} \ \ \rho \le \frac{Q\sub{max}}{V_0} 
        & \text{(free flow)}, \\
 Q\sub{max} \left[ 1-\frac{c}{V_0}\right]+c\rho
    & \quad \text{if} \ \ \rho > \frac{Q\sub{max}}{V_0}
    & \text{(congested flow).}
 \end{array}\right.
\ee
The algorithm for detecting congestions and tracking their upstream fronts is
specified completely in Chapter~8.5.8 
of~\cite{TreiberKesting-Book}, cf. also Fig.~\ref{fig:shock}. It uses
the facts that (i) in free traffic, flow 
information is propagated downstream at velocity $V_0$, (ii) in congested
traffic, the propagation is upstream at velocity $c$, (iii) the
propagation of the position $x_{12}(t)$ of the transition free $\to$
congested is given by the shock-wave formula
\be
\label{shock}
\abl{x_{12}}{t}=\frac{Q_1-Q_2}{\rho_1-\rho_2}
\ee
where $Q_1$, $\rho_1$, $Q_2$, 
and $\rho_2$ are expressed in terms of the flows of the upstream and
downstream detectors, respectively, taken at delayed
times corresponding to the propagation velocities $V_0$ and $c$, respectively. 

%#############################################################
%#############################################################

The \emph{floating-athlete data} (visualized as piecewise linear
trajectories in Fig.~\ref{fig:shock}) is used to calibrate the three
parameters $\vec{\beta}=(V_0,Q\sub{max},c)\sup{T}$ of the basic
algorithm in real time. Whenever a transition free-congested is
recognized by a new trajectory at position $x_{12}\sup{traj}(t_i)$ and
time $t_i$ (black bullets in this figure), the
parameters are updated by minimizing the objective function
\be
\label{obj}
S(\vec{\beta})=\sum_{i}
\left(x_{12}\sup{pred}(t_i)-x_{12}\sup{traj}(t_i)\right)^2.
\ee
The summation over the squared differences between the predicted and
observed jam front locations starts with the first trajectory detecting a jam
(e.g. if the speed drops consistently below a speed threshold). To
focus on recent data, an exponential weighting proportional to
$\exp[(t_i-t)/\tau]$ (where the time scale $\tau$ should contain a few
trajectories) is possible as well.

We emphasize that, even if speed data $V_i$ are available at detector $i$,
the algorithm is more robust when the densities in the denominator of
the shock-wave formula~\refkl{shock} are calculated using exclusively
the detector flows (i.e., 
the count data) and inverting the fundamental diagram rather than
directly estimating the densities by using $\rho_i=Q_i/V_i$.  

Finally, we notice that this formulation assumes that there is only one transition
free-congested between the two stationary detectors. It is most
efficient if the downstream detector is located just upstream of a known
bottleneck, and the upstream detector is just upstream of the
congested region at its maximum extension.

%#############################################################
\subsection{\label{sec:ground}Simulating the ground truth}
%#############################################################

\begin{figure}
\fig{01.00\textwidth}{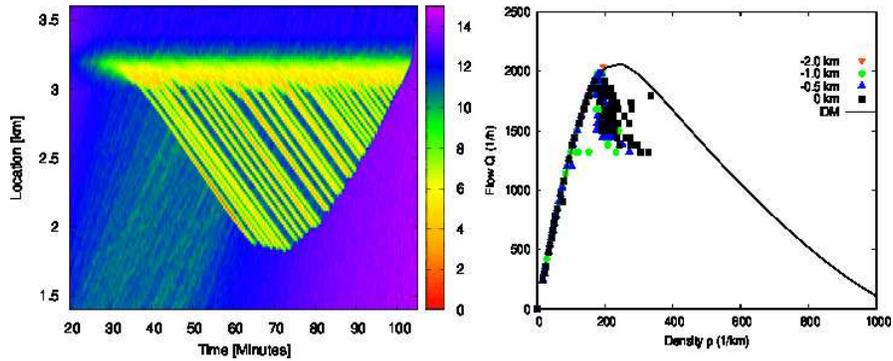}
\caption{\label{fig:groundTruth}''Virtual'' ground truth as obtained by
  an IDM simulation. Left: local density; right: flow-density data as
  obtained from stations at the indicated locations and IDM
  fundamental diagram.
}
\end{figure}

Since no true ground truth (a complete coverage of the spatio-temporal
local speed and flow) is available, the algorithm is tested by simulating
the ground truth with a \emph{completely different} model, namely the
microscopic intelligent-driver model (IDM)~\cite{Opus} to which some
acceleration noise is added to simulate heterogeneity. Particularly,
the IDM fundamental diagram is \emph{not} triangular
(Fig.~\ref{fig:groundTruth} right). Because of their
intuitive meaning, it is straightforward to adapt the IDM parameters
to directed crowd flows. Specifically, we assumed the  parameters of
Table~\ref{tab:param} representing
moderately fast Marathon runners.
%
%#############################################################
\begin{table}
\centering
\caption{IDM parameters for moderately fast Marathon runners}
\label{tab:param}       % Give a unique label
\begin{tabular}{ll}
\hline\noalign{\smallskip}
parameter & value \\
\noalign{\smallskip}\hline\noalign{\smallskip}
width in running direction $l$   & \unit[0.4]{m}  \\
maximum speed              $v_0$ & \unit[4.0]{m/s}  \\
desired time gap           $T$   & \unit[1.0]{s}  \\
minimum  gap               $s_0$ & \unit[0.5]{m}  \\
gap                        $s_1$ & \unit[0.5]{m}  \\
desired acceleration       $a$   & $\unit[0.5]{m/s^2}$  \\
comfortable deceleration   $b$   & $\unit[1.0]{m/s^2}$ \\
\noalign{\smallskip}\hline
\end{tabular}
\end{table}
%#############################################################

%#############################################################
%#############################################################

Figure~\ref{fig:groundTruth} shows the ground truth as obtained with
these parameters in an open system of variable inflow and a bottleneck
at $x=\unit[3.2]{km}$ causing the onset of congestion at about
$t=\unit[25]{min}$. Notice that the IDM is capable to produce
stop-and-go waves also for the crowd-flow parameterization. When increasing the
acceleration parameter to $a=\unit[1]{m/s^2}$, all flow instabilities
vanish.
 
%The IDM is defined by the acceleration function~\cite{Opus}
%\begin{equation}%
%	\dot{v}_{\rm IDM}(v,\Delta v,s)=a\left[1-\left(\frac{v}{v_0}\right)^4-\left(\frac{s^*(v,\Delta v)}{s}\right)^2\right]
%\label{eq:03}
%\end{equation}
%This expression combines the acceleration strategy to reach a desired speed
%$v_0$ with a braking strategy \mt{that compares the actual gap $s$
%  with the dynamically desired gap $s^*(v,\Delta v)=s_0+\max(0,vT+v \,
%  \Delta v/(2\sqrt{ab}))$}. 
%A more detailed model description can be found in \cite{Opus}.

%#############################################################
\begin{figure}
\fig{01.00\textwidth}{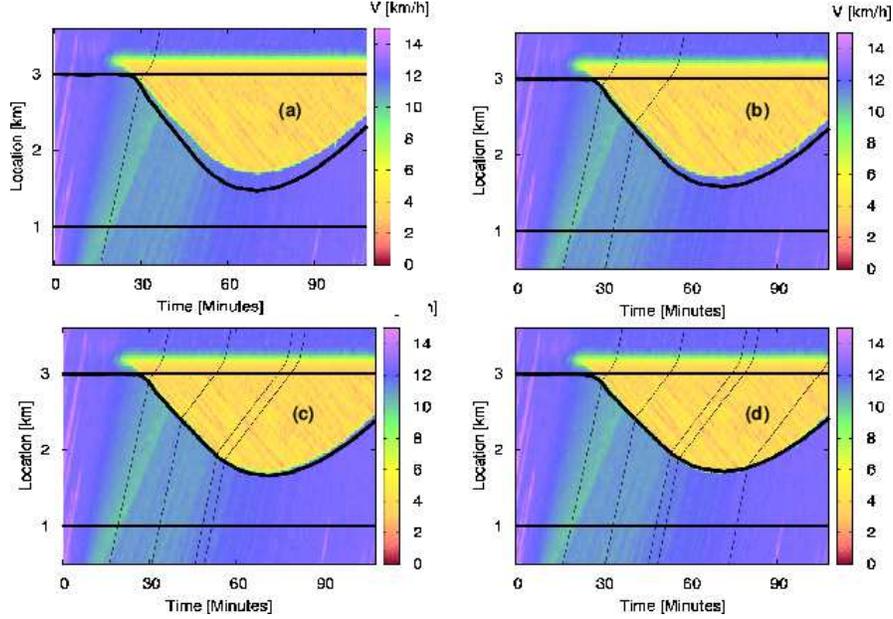}
\caption{\label{fig:dynCalib}Real-time calibration and jam front
  detection of a congestion caused by a bottleneck at
  $x=\unit[3]{km}$. The available information consists of two
  stationary detectors (horizontal lines) and ``floating-athlete''
  trajectories (dotted curves). For visualization purposes, the
  (simulated) local ground truth speed is shown color-coded in the
  background. The jam-front detection algorithm started with the
  parameters $v_0=\unit[10]{km/h}$, $c=\unit[-5]{km/h}$, and
  $Q\sub{max}=\unit[2\,300]{s^{-1}}$. (a)-(d) shows the online
  calibration when new floating-athlete trajectories become available.
  The corrected parameters where 
  (a) $v_0=\unit[9]{km/h}$, $c=\unit[-4.5]{km/h}$, $Q\sub{max}=\unit[2\,300]{s^{-1}}$, 
  (b) $v_0=\unit[9]{km/h}$, $c=\unit[-4.0]{km/h}$, $Q\sub{max}=\unit[2\,300]{s^{-1}}$,
  (c) $v_0=\unit[9]{km/h}$, $c=\unit[-4.0]{km/h}$, $Q\sub{max}=\unit[2\,350]{s^{-1}}$, 
and
  (d) $v_0=\unit[9]{km/h}$, $c=\unit[-4.2]{km/h}$, $Q\sub{max}=\unit[2\,440]{s^{-1}}$
 The black curve to the right of the last trajectory shows the jam
 front that the algorithm would predict just based on the detector
 data. 
}
\end{figure}
%#############################################################

%#############################################################
\section{\label{sec:res}Results}
%#############################################################

We test the algorithm based on the simulated ground truth generated
by the IDM as described in the previous section but with an
increased acceleration parameter $a=\unit[0.8]{m/s^2}$ resulting in
stable congested flow. The color-coded background of
Fig.~\ref{fig:dynCalib} depicts the spatio-temporal local speed which
is the same in all sub-figures (a)-(d). We assumed two stationary
detectors at $x=\unit[1]{km}$ (upstream of the congestion at any time)
and  $x=\unit[3]{km}$ (just upstream of the bottleneck at
$x=\unit[3.2]{km}$) and started the algorithm at $t=0$ (free traffic
everywhere). Sub-figure (a) depicts the predicted position of the jam
front in the past and future relative to the calibration using only
the first trajectory at $t=\unit[30]{min}$. The passage time of the jam
front at the downstream detector ($t \approx
\unit[25]{min}$) is reconstructed correctly while the congestion zone
predicted at later times is a little too extended. Subsequent
real-time calibrations after transition signals from three further
trajectories (subplots (b)-(d) gradually increase the algorithm's
quality.

%#############################################################
\begin{figure}
\fig{0.80\textwidth}{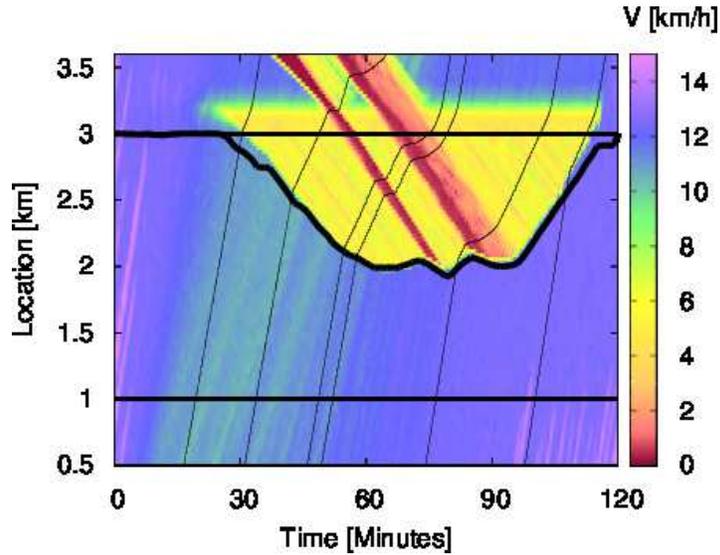}
\caption{\label{fig:valid}Validation by predicting the jam front (with
  unchanged parameters) in a
  different scenario where an additional jam propagates through the
  original jam. The floating-athlete data were not used in this test.
}
\end{figure}
%#############################################################

Finally, Fig.~\ref{fig:valid} shows a validation result by applying
the algorithm as calibrated in Fig.~\ref{fig:dynCalib}(d) to a new
situation where the congestion caused by the bottleneck at
$x=\unit[3.2]{km}$ 
is superimposed by a more severe congestion (near standstill) caused by a
further temporary bottleneck at $x=\unit[4.0]{km}$. Even when using
\emph{only the two detectors} (the trajectories are only drawn for
visual purposes), the maximum error in predicting the location of the
jam front is about \unit[200]{m} and significantly less most of the
time.

%#############################################################
\section{\label{sec:concl}Discussion}
%#############################################################
In this work, I have proposed a traffic state recognition algorithm
which is extremely fast (fractions of seconds of computation time for each
real-time calibration, much less for a prediction with fixed
parameters) and only needs sparse information: two stationary counting
detectors 
(stations where split times are taken) upstream and downstream of the
section to be analyzed, and a few ``floating-athlete'' trajectories
during the period of congestion are sufficient. Since relevant events
have of the order of 10\,000 participants, this corresponds to a
minimum penetration rate of fractions of a percent which is satisfied
in most races (cf. Fig.~\ref{fig:data} right).  By applying the algorithm
independently to each section between the two respective neighboring stations, the
whole course can be covered. Application of the calibrated algorithm
to a new situation results in comparatively low errors demonstrating
its robustness.

It is worth noticing that the algorithm ignores all lateral
dynamics, i.e., it is based on only the longitudinal dimension. Among
others, this allows using car-following models for generating the
ground truth although they represent single-file motion rather than
true crowd dynamics. The rationale for that is the inherent
competitiveness of the athletes always seeking the best lateral
position (there is no ``go right directive'' in racing events) and
thereby leveling off lateral differences of the longitudinal speed
(i.e., shear rates).

There is evidence that the discrepancy between the free-flow regime of
the triangular LWR and the actual free-flow dynamics of the crowd (or
the IDM) contributes
significantly to the residual errors. This can be improved if the
LWR free-flow dynamics is replaced by the 
dispersion-transport model presented at TGF'13~\cite{TGF13-running}.

In future investigations, the proposed algorithm will be tested using
more realistic two-dimensional microscopic crowd flow models to
generate the
ground truth. It is also planned to test it against the incomplete
ground truth obtained from data of real mass-sports events. Finally,
we will apply this algorithm to vehicular traffic flow. 

%\bibliographystyle{elsart-num}
%\bibliography{database}

\input{Treiber_datafusion_small.bbl}
\end{document}

%% file: defsSubmit.tex
\usepackage{units} %\unit[3]{m/s^2} => 3 m/s^2 etc!

\providecommand{\bc}{\begin{center}}
\providecommand{\ec}{\end{center}}
\providecommand{\be}{\begin{equation}}
\providecommand{\ee}{\end{equation}}
\providecommand{\bea}{\begin{eqnarray}}
\providecommand{\eea}{\end{eqnarray}}
\providecommand{\bdm}{\begin{displaymath}}
\providecommand{\edm}{\end{displaymath}}
\providecommand{\bdma}{\begin{eqnarray*}}
\providecommand{\edma}{\end{eqnarray*}}
\providecommand{\ba}{\begin{eqnarray*}}
\providecommand{\ea}{\end{eqnarray*}}
\providecommand{\bi}{\begin{itemize}}
\providecommand{\ei}{\end{itemize}}
\providecommand{\benum}{\begin{enumerate}}
\providecommand{\eenum}{\end{enumerate}}

\providecommand{\refkl}[1]{(\ref{#1})}

\providecommand{\text}[1]{{\mbox{ #1}}}

\providecommand{\fig}[2]{
   \begin{center}
     \includegraphics[width=#1]{#2}
   \end{center}
}

\providecommand{\abl}[2]{\frac{{\rm d} #1}{{\rm d} #2}}  %d(#1)/d(#2)

\providecommand{\sub}[1]{_{\rm #1}}
\renewcommand{\sup}[1]{^{\rm #1}}